\documentclass[final,12pt]{elsarticlemod}

\usepackage{graphicx}
\usepackage{amsmath}
\usepackage{amssymb}
\usepackage{lineno}
\usepackage{url}
\usepackage{hyperref}
\usepackage{multirow}

\usepackage[all]{nowidow}
\expandafter\def\expandafter\UrlBreaks\expandafter{\UrlBreaks\do-}

\newlength{\mylen}
\setbox1=\hbox{$\bullet$}\setbox2=\hbox{\tiny$\bullet$}
\journal{Tokenomics 2019}
\date{May 6, 2019}

\begin{document}

\begin{frontmatter}


\title{
    Democratising blockchain:\\ A minimal agency consensus model
    }




\author[email-marcin]{Marcin Abram}
\author[email-david]{David Galindo}
\author[email-daniel]{Daniel Honerkamp}
\author[email-jon]{Jonathan Ward}
\author[email-jin]{Jin-Mann Wong}
\address{Fetch.AI, St.\ John's Innovation Centre, Cowley Road, Cambridge CB4 0WS, UK}

\fntext[email-marcin]{marcin.abram@fetch.ai}
\fntext[email-david]{david.galindo@fetch.ai}
\fntext[email-daniel]{daniel.honerkamp@fetch.ai}
\fntext[email-jon]{jonathan.ward@fetch.ai}
\fntext[email-jin]{jinmann.wong@fetch.ai}  

\begin{abstract}
We propose a novel consensus protocol based on a hybrid approach, that combines a directed acyclic graph (DAG) and a classical chain of blocks. This architecture allows us to enforce collective block construction,
minimising the monopolistic power of the round-leader.
In this way, we decrease the possibility for collusion among senders and miners, as well as miners themselves, allowing the use of more incentive compatible and fair pricing strategies. We investigate these possibilities alongside the ability to use the DAG structure to minimise the risk of transaction censoring.
We conclude by providing preliminary benchmarks of our protocol and by exploring further research directions.
\end{abstract}


\end{frontmatter}

\nolinenumbers



\section{Introduction}
\label{S:1}

Most popular consensus protocols, such as Proof of Work (PoW) \cite{Nakamoto2008, satoshi_literature} and Proof of Stake (PoS) \cite{king2012ppcoin, on_stake} resemble a rent-seeking contest \cite{Tullock1980, Thum2018} for the right to create the next block. The contest usually consists of selecting a participant from a population of miners, proportional to a costly sacrifice they have made, such as hashrate in PoW or stake in PoS. The winner gains a monopoly to create the next block, meaning they are free to compose a block of any available transactions they wish, within certain protocol limits such as block size \cite{mastering-bitcoin}.

The high dependence of PoW mining on energy costs has led to a geographical concentration of miners in areas where energy is cheap, such as China \cite{kaiser2018looming}, while economies of scale and a high variance in rewards has led to a further concentration of power in the hands of operators of mining pools \cite{Gervais2014, Beikverdi2015, Arnosti2018}.
With this inherent progression towards centralisation there is a rising threat that some users or transactions can be censored. This issue represents a threat to ``freedom of speech'' in the context of on-chain governance mechanisms and a security threat in the form of punitive and feather forking attacks \cite{Bonneau2015, Mosakheil2018}. In addition selective transaction censoring can also lead to manipulation of and security threats to financial protocols such as decentralized exchanges \cite{censorship}.

As the security of a decentralised blockchain depends on the total mining rewards (transaction fees and block rewards) \cite{BiS, budish}, the ability to extract fees is essential to the system. 
Allowing users to freely decide on the fee attached to their transaction is the current most popular approach. Given each individual miner's incentive to maximise the fees they can extract with their block, this approach relies on restricting throughput on a protocol level to generate fees \cite{Huberman2017, Lavi2017}. If block producers act as profit maximisers and select transactions according to their fee, the most prevalent fee selection mechanism (used by, among others, Ethereum and Bitcoin) can be characterised as a generalized first price auction \cite{Nakamoto2008, Lavi2017, ethereum}. While numerous other mechanisms have been proposed, the monopoly of the block producer to select transactions heavily constrains the mechanisms that can be employed: unless the block producer is awarded a hundred percent of the fees, the producer can always gain by circumventing the protocol. Colluding with a transaction sender through a third channel payment, the sender can simply pay the block producer the original fee $f - \epsilon$ while sending the transaction with the lowest possible fee $\epsilon$ to the network.
On the other side of the spectrum we can find algorithmic fee setting mechanisms. The main difficulty with these approaches is that they have to internalise the externalities a transaction is causing, which include system security, usage of computation, bandwidth and storage as well as decentralisation.

The time propagation of blocks, for sizes larger than 20kB, is almost linear with their size \cite{Decker2013}. It can therefore be more profitable to mine empty blocks in situations where the block reward is much larger than the sum of the transaction fees and if there is a competition between soft forks, such as in Bitcoin \cite{kaiser2018looming, propagation-cost}. This dynamic lowers the efficiency of the system and should therefore be mitigated against in protocols aiming to maximise throughput. One method to address this issue, as well as the censorship, could be to increase the cost of creating a sub-optimal block to the producer by scaling their rewards by the proportion of transactions included in their block.

Another problem with the incentive structure of many blockchain systems, such as Bitcoin and Ethereum, is the lack of incentive for nodes to propagate transactions \cite{Babaioff2011}. Information sharing is carried out by the nodes in these networks, which may or may not additionally participate in mining, and is not rewarded. This is therefore only sustainable when the cost of running a node is sufficiently low so that the system is able to maintain a large population of nodes relative to miners. In directed acyclic graph (DAG) protocols the dynamic is slightly different as nodes have an incentive to gossip transactions that build upon their own \cite{iota, byteball}. 

The problems of censorship and limited fee pricing algorithms are to a large part rooted in the monopoly over block construction. Here we will propose a new way to collaboratively pack a block through the use of a DAG, and to then discuss the implications and benefits of minimising the block proposer's freedom.

\section{Literature review}
\label{S:2}

The Prism consensus protocol \cite{Prism} aims to maximise the transaction throughput of a decentralised payment system by parallelising the task of collecting transactions, producing blocks and reaching finality on the blockchain. Although the idea of separating these processes has appeared in other protocols \cite{fruitchains, algorand, inclusive-blockchains}, very few completely divide all three. The Prism proposal shares many similarities with our model described in Section \ref{S:3}. In particular, they also maximise the throughput of the system by incentivising the sharing of transactions sets, in their case in the form of transaction blocks, alongside the block production, allowing block producers to propagate smaller final blocks. However, their protocol does not use this to limit the censorship power of the block producer, who is still free to create an empty block or a block that contains only transactions of their own choice.

The model described in the next section builds upon protocols utilising a decentralised round based random seed, which has been implemented in many protocols \cite{ouroboros, algorand, dfinity}, through the introduction of a DAG. Here we focus on the DFinity protocol which makes use of a verifiable random function (VRF) \cite{VRF} to enable the notarisation of blocks by a committee in each round. The system is designed to accommodate high throughput while being resistant to various common attack vectors. Double-spend attacks involving the withholding blocks, such as Finney attacks \cite{finney}, are defended against, in these systems, by requiring the notarisation of each block by a random committee in each round. The difficulty of such an attack is therefore increased by the attacker now being being required to control 51\% of the committee, and having the highest-ranking block proposal. The protocol also detects attacks involving network partitions and resolves these by either pausing the protocol, preventing further blocks from being generated, or only continuing on the majority branch \cite{dfinity}.

The consensus protocols of IOTA \cite{iota} and Byteball \cite{byteball} also make use of a DAG as the structure naturally scales to accommodate a high transaction throughput. In these protocols, users append their own transactions onto the graph and in the process validate the transactions to which their transaction is attached. In this way, IOTA is able to offer zero nominal transaction fees\footnote{However, note that to submit transactions IOTA users must solve a PoW problem, which is a cost on the transactor and serves a similar purpose to fees.} and the security of the system is maintained directly by its users. However, despite its advantages, the DAG only defines a partial ordering of transactions and therefore additional layers in the protocol are required to establish a strict ordering of transactions. The strict ordering is necessary to implement smart contracts and both those protocols rely on trusting a third party to do so.

In the next section we show that by using the DAG, in combination with the traditional chain structure, we can obtain a strict ordering of transactions without relying on trusted parties.

\section{Minimal agency consensus}
\label{S:3}

We propose a model that constrains the monopolistic power of a block proposer. 
The fundamental data structure that is used to record transaction events is a DAG, that we denote as $G=(V, E)$, where $v_i \in V$ are the vertices that contain collections of transaction hashes and $E$ represents the set of edges. In this model each vertex has two out-going edges\footnote{In general, the vertex out-degree $k_{\mathrm{out}}\ge 2$. The lower bound is chosen to minimise additional storage requirements.}, which are hash-references to previous vertices. The tips of the DAG refer to vertices with no in-coming edges. A DAG provides partial ordering in the sense that the children of $v_i$, the set of vertices that are both directly and indirectly referenced by a specific vertex, $v_i$, occurred with certainty in its past.

The ledger uses a decentralised random beacon (DRB) as a mechanism to elect a committee $\mathcal{C}$ that collectively agree on the next block to be published, similarly to other protocols that use a DRB \cite{dfinity}. The value of the random beacon, $s_r$, for the current round is also used to define a ranking of staked nodes in the network which specifies their priority for being the publisher of a block in the current round. We refer to members of this group as vertex-proposers. Furthermore, in our model the random beacon selects a set $\mathcal{A}$ of vertex-attachers. Each attacher has the right to append one vertex to the DAG per round, containing a list of hash-references to all transactions not yet included in any vertex in its past.
At the end of a block round, the highest ranked vertex-proposers can propose a number of vertices (and thereby transactions) to be packed into the next block, which are then notarised by the committee based on the proposer's rank. In particular, each vertex-proposer signs the following information: a proposed set of vertices, previous block hash, and the hash of the Merkle tree of transactions in the block.
The signed proposals are propagated around the network and all committee members notarise the highest ranked proposals they receive. Under certain assumptions, the protocol can then guarantee finalisation of a notarised proposal after two rounds, based on the overall ranking of the blocks in the different chains \cite{dfinity}. Once a proposal is finalised, each participant can create the block of packed transactions, defined from the proposed vertices, and can discard that part of the DAG structure that was included in the block, creating very low additional memory requirements for participants.
Given the chosen vertices, further protocol-level constraints, e.g. on price or block-size can then be imposed on the transaction if desired. Transactions covered by the vertex-proposal that were not included in the block, can then be carried over to the next round, be included again in new vertices or rejected (leaving it to the sender to re-send them).

To maximise system throughput with the DAG, the protocol has to incentivise: 
\begin{itemize}
\setlength\itemsep{0em}
    \item[i)] attachers to build a compact DAG, which allows the proposers to maximise the descendants of any proposed set of vertices as much as possible
    \item[ii)] the proposers to select the vertices maximising the number of descendants
\end{itemize}
The minimal agency of the vertex-proposer is enforced by the collaborative construction of the DAG, which links together groups of transactions, and in incentivising vertex selection by ii). This latter constraint can be imposed in various ways, each with a different cost of censorship to the vertex-proposer, which are discussed in section \ref{S:4}. Note that, given honest actors, ii) should be almost equivalent to selection by fee maximisation (with difference only stemming from some participants not yet being aware of certain transactions).

\section{Results}
\label{S:4}

In this section we detail previously unenforceable strategies that can be employed with our model and how they can  overcome the shortcomings outlined in section \ref{S:1}. 

\subsection{Fees}\label{S:fees}

Fees in decentralised blockchain systems have two major purposes: first, to secure the system by encouraging sufficient mining activity for the cost of an attack to be high enough to deter attacks, and second, to distribute new tokens. Here we are concerned with the ability to efficiently extract fees to provide the necessary security. In this section we will deal with the more difficult scenario where additional, inflation-financed, block rewards are not present. Several of the most popular cryptocurrencies, including Bitcoin and Ethereum~1.0, are based on a fixed total token supply, meaning any inflation can only be temporary and will eventually need to approach zero. Unless the system operates at its absolute transaction limit, the externalities between a miner's incentive to maximise their block reward and the social good of system security\footnote{Further externalities stem from the requirement that all (future) miners will have to compute and store the transaction, but the reward is only attributed to the block producer.} imply that we cannot let the free market decide on both block size and price of transactions \cite{freemarket-fees}. To account for these inefficiencies, the protocol has to impose some form of constraint by either limiting the miner's ability to choose a block size or the users ability to set a price.

The prevailing approach, used in protocols such as Bitcoin and Ethereum, is to set an explicit absolute limit on the block size and let the users freely bid for a place in a block, resulting in what is effectively a first price auction\footnote{In practice, users are motivated by the time required for the transaction to be entered into a block. This wider competition for space in block's that are being produced at a constant rate has been shown to be a Vickery-Clark-Groves mechanism}. The literature around this pricing mechanism and its application in blockchain technology has identified a number of weaknesses and alternative mechanisms. However, a block producer's freedom to select transactions combined with the existence of third channel payments renders many of these mechanisms unenforceable. We will now discuss some of the most common ones and show how our proposal clearly extends the set of pricing mechanisms that can be employed.

\textit{Decoupling mining rewards:} when the Bitcoin protocol fully transitions to transaction fee-based block rewards, miners may no longer be incentivised to mine on the longest chain. Intuitively, this arises when the shorter alternative chain includes blocks that contain lower total transaction fees. The availability of a larger number of high-value transactions means that mining a block on the shorter chain can lead to a higher reward for the miner. This can decrease chain stability due to increased forking, lead to equilibria with partially empty blocks and increase the gains from selfish-mining strategies \cite{decouple-block-reward}. By decoupling miners' rewards from their blocks, this misalignment of incentives can be removed. This idea has been explored by a few cryptocurencies, such as Ouroboros \cite{ouroboros} and Fruitchains \cite{fruitchains}. The idea here is that miners are instead rewarded based on the block rewards of a sequence of $Q$ blocks. While this solves the outlined problem by making them indifferent as to which block contains a particular transaction (within the defined window $Q$), it suffers from two other issues: first, it reduces the marginal return of a miner with regard to a specific transaction, reducing the incentive to include the highest-paying transactions and to build blocks with the highest possible reward (with $Q$ approaching infinity the miner would become increasingly indifferent between the alternatives of creating an empty and a block full of transactions). Furthermore, the block producer can now profit from colluding with transaction senders: instead of sending a transaction with a fee $f$, the sender can directly pay the producer $f-\epsilon$ and send the transaction to the network with a negligible fee $\epsilon$. As the proposer cannot freely choose the transactions in the block, this type of indirect payment cannot be prevented.

\textit{Share mining rewards between different functions:} Many protocols have developed more complex differentiation between different network functions, such as block proposing and voting \cite{Prism}, or including off-chain blocks \cite{inclusive-blockchains, GHOST}. In many cases, it would be desirable to directly link these functions to the block fees and in the absence of inflation based block rewards this is a necessity. However, by awarding non-block producing roles with a share $x$ of the rewards, we reduce the reward of the producer to $(1-x)$ of the full block reward. This approach reduces the incentive for including transactions in a block, similarly to the decoupling approach, and leads to the same outcome where a block producer can gain by colluding with a transaction sender through third-channel payments.

\textit{$k^{\mbox{\scriptsize th}}$ price auctions:} first price auctions require complex and inefficient strategies to deduce a good bid \cite{auction-anarchy}. A classical alternative are $k^{\mbox{\scriptsize th}}$ price auctions: as a higher bid does not marginally affect the resulting fee, the simple optimal strategy is to bid one's true willingness to pay \cite{Lavi2017}. The reason this has not been implemented in existing blockchains is its vulnerability to collusion: a block producer can bribe a transaction sender, who is willing to pay a fee $f$ to instead send their transaction with a fee $f + \Delta$ and privately refund them with the bribe $\Delta$. This raises the price for everyone else while still reaping fees from the low paying transactions\footnote{Even simpler, excluding some low-fee transactions, the block producers could simply include dummy transactions to themselves with high transaction fees to raise the price level to a (for themselves) more optimal level.} \cite{BlockchainResourcePricing}. While this strategy does not depend on the block producer's monopoly, it quickly becomes costly if the block producer does not receive 100\% of the transaction fees of the block. This in turn depends on limiting the producers freedom, as outlined above.

\textit{Algorithmic pricing:} the alternative to limiting the block size and allowing users decide the transaction fee they bid is to instead prescribe an algorithm on the protocol level that determines the price of a given transaction. While some discussions of this exist \cite{BlockchainResourcePricing, Prism}, we are not aware of any implementations. The main difficulty stems from finding a price that reflects the externalities a transaction imposes on the system. These externalities include, but is not limited to: security, processing and bandwidth cost for all nodes currently online, and the long-term storage cost of maintaining the state for every node. Crucially, due to these externalities, the optimal price depends not only on characteristics of the transaction itself (such as byte size or in the case of a smart contract complexity), but on all the other transactions submitted to the system. While this direction is promising, it still requires further research. A potential benefit of our proposed protocol could be that, by first establishing consensus on collectively added transactions, all participants essentially create a ``common memory pool", on top of which pricing algorithms could be applied. This might enable the algorithm to take into account statistics that would otherwise only exist outside of the protocol, such as the current demand and congestion.

\textit{Empty blocks:} In situations where (inflation-financed) block rewards make up the largest part of the block reward, the race between block producers, and the resulting risk of a block becoming an orphan, can make it more profitable to publish smaller blocks that propagate faster than full blocks that include the additional transaction fees \cite{kaiser2018looming, propagation-cost}. While this maximises the miner's rewards, it is obviously socially undesirable. By constraining the proposer's ability to select the transactions and decoupling the attachers' incentives from the propagation time of the block, this behaviour can be made unprofitable.

\subsection{Censorship}
If we assume that the attachers of vertices have a different identity to the proposers or notarizers of the DAG, the exclusion of a specific transaction requires the proposer to forgo the fees of all vertices that build upon vertices containing this transaction. However, anonymity means that a participant can create multiple identities and therefore participate in both groups or, in absence of this, attachers and proposers could collude to attach alternative vertices including all but the censored transaction\footnote{An alternative would be to restrict the size of a vertex, but avoiding hard-coded rules is generally desirable.}. This would allow the proposer to simply ignore the vertices appended by other attachers, bringing us back to a monopoly over the next block.

To prevent the this situation from occurring, and to make the proposals a fair representation of the collectively built vertices, it would make sense to provide incentives to force proposers to maximise the number of descendants of the selected vertices\footnote{In protocols where each identity represents a fixed stake, as required for many VRFs, this coincides with the maximisation of the total stake associated with the DAG.}. As noted earlier, this is still almost equivalent to maximising the fee covered, but strips away the proposer's power to decide alone on the block content.

Given a global view of the DAG, this objective can easily be enforced. However, abandoning the synchronicity assumption, this becomes non-trivial as we cannot ensure that every participant will reach the same conclusion about the optimal decision within a finite time. Nonetheless, we can implement approximations of this selection rule. In doing so, we differentiate between soft (incentive based) and hard constraints:

\begin{itemize}
    \item[a)] Hard constraints: impose a function $f(n_{vertices}, x)$ that any proposal has to satisfy, where $n_{vertices} = |\mathcal{A}|$, i.e. the theoretical maximum number of vertices that could be covered. The main difficulty in this approach is that the effective maximum in a given round varies with the structure of the DAG in that round. The function $f$ can either be a simple requirement of the number of descendants to be a minimum percentage of vertices or a more complex approximation of the effective maximum. Critically, it can only depend on further inputs $x$ that are verifiable even with a partial view of the DAG. While a very promising direction, defining a robust rule with guaranteed and non-interactive resolution under all circumstances will require further research. 
    \item[b)] Soft constraint: reward the proposer proportionate to $$\delta = \frac{n_{descendants}}{n_{vertices}}\,,$$ where again $n_{vertices} = |\mathcal{A}|$, and $n_{descendants}$ is the number of descendants of the proposal. The cost of circumventing the DAG will quickly grow proportional to the depth of any vertex the proposer wants to ignore. Note that this introduces an element that is outside of the proposer's control, as the best $\delta$ achievable in a given round depends on both the attacher and the network delays that occurred in this round. This does come at the cost of increased volatility of an individual proposer's rewards. However, the possibility to now more broadly distribute the rewards can still result in an overall lower volatility, when taking into consideration the different roles of a staker.
\end{itemize}
While both types of constraints ensure that a profit-maximising agent will choose to maximise the number of descendants and thereby enable the advantages outlined in section \ref{S:fees}, only the hard constraint makes it impossible to ignore specific vertices within the DAG. Since under the soft constraint it is possible that an attacker's gain from censorship will be larger than the economic loss of the rewards from the round.

A third alternative would be a competitive protocol, which optimises the throughput of the network, where the vertex-proposals are ranked according to a function of the proposer's ranking and the number of descendants. This encourages proposers to make maximal use of their knowledge of the DAG for fear of losing out on the block reward.

\subsection{Efficiency and Throughput}

Most existing blockchains, including Ethereum and Bitcoin, combine the role of adding new transactions and defining the order between them. This coupling is a major limitation of both block size and the block production rate. Increasing either property leads to an increase in the forking rate and thereby reduces the proportion of byzantine participants $\beta$ that the protocol can tolerate. These relationships and tradeoffs have been studied extensively for the Bitcoin protocol \cite{GHOST, bitcoin-security-synchronous, bitcoin-security-asynchronous}. While more recent protocols, such as GHOST \cite{GHOST}, have altered the fork choice rule to incorporate ``uncle blocks", a similar tradeoff between the block production rate and $\beta$ exists. In this case the system becomes vulnerable to balancing attacks \cite{balancing-attack}, in which an attacker splits the work of honest nodes into equal subtrees.
By decoupling the addition of new transactions from block production (delaying specification of the final ordering), this constraint can be lifted, which allows transactions to be added much more rapidly. Switching from a stochastic block production rate to comparably deterministic block times dictated by the random beacon further alleviates this tradeoff. The nature of the DAG then allows the communication of proposals for the set of vertices to be propagated quickly, as they imply all vertices in their past.

The reduction in bandwidth from this can be calculated as follows: let us assume we would like to achieve $n_{tps}$ transactions per second, and the block time is $t_{block}$. At a bare minimum this requires the propagation of vertices, assuming each 32 byte transaction hash appears in the DAG only once, during each round involves
$$
    B_{\text{DAG}} = (32 n_{tps} t_{block} + 129 n_{vertices}) \text{ bytes} \,,
$$
where the coefficient in front of $n_{vertices}$ contains the size of a signature and two transaction hashes, which are the links for the vertex. A compact block\footnote{A compact block introduced in Bitcoin \cite{compactblock} replaces the list of transactions in a block with a list of 6 byte hashes, vastly reducing the data transmitted through the network. The idea utilises the fact that the memory pools between nodes generally contain the same transactions and therefore propagating full transactions in blocks is mostly redundant. In this scheme, any node that does not possess the transaction corresponding to a hash in the compact block, can simply query its peers for the information.}, ignoring the block header, containing the same set of transactions has size $6 n_{tps} t_{block}$ bytes. Assuming that there are an $O(1)$ number of block proposals each round, and neglecting the size of tip proposals, the gains in terms of bandwidth from the DAG are of $O(1)$ (meaning, that it is constant and does not depend on the number of transactions) and come from the fact that $B_{\text{DAG}}$ is not sent through the network in a single event but is synchronised continuously over the entire period of the round.
    
As mentioned in section \ref{S:3}, in order to maximise the throughput of the system, the protocol should incentivise the vertex-attachers to create a compact DAG from which the vertex-proposer can easily maximise the number of descendants. The choice of committee members for vertex-attachers allows for various incentivisation schemes depending on whether this group consists of all users or is a random subset. In the former case, the incentives align with those of IOTA and Byteball where vertex-attachers are naturally incentivised to attach vertices to the DAG without a reward (as they will contain their own transactions), and they will choose to implement the algorithm that maximises the probability of their transactions being included in the next block. In the latter case, to avoid a tragedy-of-the-commons situation and to maximise the transaction throughput it is necessary to either reward the vertex-attachers, or punish i.e.\ through stake slashing \cite{slasher}, the nodes that do not cooperate.

In order to evaluate the effectiveness of these strategies, we simulated the construction of a DAG using various algorithms to append vertices, varying also the size of the group allowed to attach vertices to the DAG. Four different algorithms were studied:
\renewcommand\labelitemi{\raisebox{\mylen}{\tiny$\bullet$}}
\begin{itemize}
\setlength\itemsep{0em}
    \item \emph{Random}: the tips are chosen randomly from available orphans.
    \item \emph{Joint cardinality}: the pair of tips which have the maximum joint span of vertices, also referred to as the cardinality, is chosen. More concretely, the union of the two sets of descendants is taken and its size considered.
    \item \emph{Metropolis}: pairs of tips are chosen at random until the joint cardinality of the tips is within the threshold of the total number of vertices.
    \item \emph{Greedy}: links are chosen by maximising cardinality on each link separately. First one tip is selected, and then the second is selected by considering the cardinalities of the remaining tips. This differs from the joint cardinality algorithm in that the set of descendants of each tip are considered separately even though the same elements may appear in both. 
\end{itemize}

\begin{table}
\centering
\begin{tabular}{c|c|c|c}
\multirow{2}{*}{Algorithm Type} & \multicolumn{3}{c}{Vertex Proposal Size} \\
& $n_{vertices} = 10$        & $n_{vertices} = 100$       & $n_{vertices} = 1000$       \\\hline
\emph{Random} & 2           & 15          & 144          \\\hline
\emph{Joint Cardinality} & 2           & 24          & 308          \\\hline
\emph{Greedy} & 2 &  20 & 209 \\ \hline
\emph{Metropolis} & 2 & 15 & 145 \\ \hline
\end{tabular}
\caption{The average size of a vertex proposal across 100 blocks for each DAG construction algorithm for $n_{vertices} = 10,\ 100,\ 1000$. In this simulation tips of the DAG are discarded if they are older than 10 block rounds.}
\label{tab:metrics}
\end{table}

The size of the vertex proposals required to capture all vertices appended in the previous round are shown in table \ref{tab:metrics} and shows that allowing a large committee of vertex-attachers significantly increases the size of vertex-proposals that are needed to specify the DAG. It is therefore more beneficial for the efficiency of the system to restrict vertex-attachers to a small fraction of staked nodes to balance security of the system with the feasibility of vertex proposal by maximising the number of descendants. 

\section{Conclusions}

In this article we have proposed a new DAG-based consensus protocol which leverages the separation of functions to both minimise agency and maximise throughput.

We describe incentive incompatibilities and restrictions stemming from the monopolistic power of a block producer to select transactions, once they have won the right to produce the next block. By restricting their freedom to do so, we eliminate issues that prevent existing protocols from implementing a range of fee pricing mechanisms. We further discuss different approaches to incentivise or enforce collective block construction in asynchronous networks. While more research is needed to proof the security and efficiency of these new approaches, the strict increase in enforceable pricing schemes opens up new opportunities to extract the fees necessary to secure a decentralised system in the absence of inflation-financed block rewards.


\bibliographystyle{bib-style}
\bibliography{bibliography.bib}

\end{document}